# On LiF:Mg,Cu,P and LiF:Mg,Ti phosphors high & ultra-high dose features

Barbara Obryk[a*], Helen J. Khoury[b], Vinicius S. de Barros[b], Pedro L. Guzzo[b], Paweł Bilski[a]

[a]Institute of Nuclear Physics (IFJ), ul. Radzikowskiego 152, 31-342 Kraków, Poland

[b]Federal University of Pernambuco (UFPE), Av. Prof. Luiz Freire 1000, 50740-540 CDU Recife, PE, Brazil

**HIGHLIGHTS**

• TL peak 'B' occurs for LiF:Mg,Cu,P after high doses of all radiation types.
• Highly irradiated LiF:Mg,Ti glow-curve shape depends on the radiation quality.
• High dose features of both phosphors are significantly different in many aspects.
• Associated PL/TL high-dose measurements are possible using LiF:Mg,Cu,P.
• Dopants role is crucial for high-dose features of lithium fluoride based phosphors.

**Abstract**

LiF:Mg,Ti and LiF:Mg,Cu,P are well known thermoluminescence (TL) dosimetry materials since many years. A few years ago their properties seemed well known and it was widely believed that they are not suitable for the measurement of doses above the saturation level of the TL signal, which for both materials occur at about 1 kGy. The high-dose high-temperature TL emission of LiF:Mg,Cu,P observed at the IFJ in 2006, which above 30 kGy takes the form of the so-called TL peak 'B', opened the way to use this material for measuring the dose in the high and ultra-high range, in particular for the monitoring of ionizing radiation around the essential electronic elements of high-energy accelerators, also fission and fusion facilities, as well as for emergency dosimetry. This discovery initiated studies of high and ultra-high dose characteristics of both these phosphors, which turned out to be significantly different in many aspects. These studies not only strive to refine the method for measuring high doses based on the observed phenomenon, but also, and perhaps above all, bring us closer to understanding its origin and essence. This manuscript aims to review existing research data on the high and ultra-high dose features of both LiF based phosphors.

Keywords: Thermoluminescence; Lithium fluoride; High-dose high-temperature TL emission; Peak 'B'; High-level dosimetry

* Corresponding author: e-mail: barbara.obryk@ifj.edu.pl; tel.: +48-12-6628280; fax: +48-12-6628066

## 1. Introduction

A big challenge of dosimetry nowadays is the growing demand for high-level dosimetry materials and methods due to intensive development of radiation technologies (among them materials testing, sterilization and processing) and nuclear installations (high-energy accelerators for research, e.g. Large Hadron Collider - LHC, and hadron therapy of cancer, also fission and fusion power facilities, e.g. International Thermonuclear Experimental Reactor - ITER) including accident dosimetry at these facilities (Bilski et al., 2007a). In connection with these growing needs the development of high-level dosimetry methods occurred and still is progressing (Schönbacher et al., 2009). Their number is very large, some are more universal, but most is useful in some special applications (Benny and Bhatt, 2002; Göksu et al., 1989; McLaughlin, 1996; Teixeira and Caldas, 2012; Wieser and Regulla, 1989). Among the most popular passive systems usable at high doses are polymer-alanine dosemeters, radio-photoluminescence glass detectors, optical absorption dosemeters (e.g. LiF single crystals, dyed polymeric foils), liquid chemical dosemeters (e.g. Fricke dosemeter) and hydrogen pressure detectors. In addition calorimetry and ionisation chambers are still used; however, new developed active systems are based on semiconductor p-FET or MOS-FET detectors (field effect transistors based on silicon dioxide), PIN diodes (p-n junction with built-in intrinsic undoped layer), also optically stimulated luminescence and diamond based detectors.

It is worth mentioning that none of the known high-level dosimetry materials and methods include both the dose range typical for radiation protection (miligrays) and high-dose range (kilograys). Surprisingly, to meet this challenge, came forth the discovery of unexpected properties of highly sensitive LiF:Mg,Cu,P phosphor at high doses and high temperatures (Bilski et al., 2008b; Obryk et al., 2009, Obryk, 2010). It was very fortuitous and has enabled the development of a method of measurement in the dose range of twelve orders of magnitude with a single thermoluminescent (TL) detector (Obryk et al., 2011a; Obryk, 2010, 2013). Recently observed features of this highly sensitive TL material inspired further research on high and ultra-high dose characteristics of both lithium fluoride based phosphors routinely used for TL dosimetry: LiF:Mg,Cu,P and LiF:Mg,Ti.

## 2. Materials and current status of research

### 2.1. LiF based phosphors general features

Lithium fluoride based phosphors have long been well known luminescent dosimetric materials (e.g. McKeever et al., 1995; Vij, 1993). LiF:Mg,Ti (MTS, equivalent to TLD-100) is used for the production of thermoluminescent detectors widely applied in dosimetry of ionizing radiation from the 1960s by the dosimetry services worldwide (e.g. Stadtmann et al., 2011; Obryk et al., 2011c). LiF:Mg,Cu,P (MCP, equivalent to TLD100H) was introduced in 1980s (Nakajima et al., 1978), and due to its high sensitivity is now equally widely used in TL dosimetry, especially for environmental measurements (e.g. Budzanowski et al., 2004; Ilgner et al., 2010). Main dopant concentration (related to trapping centres), magnesium for both materials, is ten times higher for MCPs (about 0.2 weight %) than for MTS, while the role of luminescence activator is played by titanium for MTS (10-15 ppm) and by phosphorus for MCP (1-4 weight %); in addition copper (0.02-0.05 weight %) for MCP plays a role still not completely clear, also the oxygen impurities seem to be important for MCP (Bilski, 2002; Chen and Stoebe, 1998; 2002). The glow-curve structure for both materials is quite similar, the main dosimetric peak (peak 5 for MTS while peak 4 for MCP) occurs at a temperature of about 220°C, preceded by a few minor peaks. The main peak's activation energy and frequency factor are similar for both materials ($E>2$ eV; $s>10^{20}$ s$^{-1}$), which implies similarity of TL processes related to both main glow peaks of these phosphors (Bilski, 2002; Horowitz, 1993). Both phosphors emit light in the short-wavelength part of the spectrum (McKeever et al., 1995), but emission bands differ, being in the range 420-460 nm for MTS while ca. 380 nm for MCP, which TL spectral characteristic have been thoroughly investigated by McKeever in 1991.

The most apparent difference between both phosphors is in their sensitivity to radiation: approximately thirty times higher for LiF:Mg,Cu,P than for LiF:Mg,Ti for gamma radiation (Horowitz, 1993; McKeever et al., 1995; Bos, 2001; Bilski, 2002); taking into account the high MTS internal background (i.e. TL signal of unexposed detector) the ratio grows to the level of about hundred. The detection threshold of MCP detectors is below 1 μGy while for MTS in the range of 20-50 μGy. Linearity range for both materials end up at a few Gy, while the upper limit of a useful dose range (manifested by a decrease of dose sensitivity to an unacceptable value), i.e. saturation dose is about 1 kGy. Another important difference between the dosimetric properties of these phosphors is in their dose response. MTS features the well-known linear-supralinear response, while MCP dose response is linear-sublinear. The sublinear dose response of MCP has some further consequences. It is generally accepted that this is responsible for the much lower TL efficiency with which heavy charged particles and high-LET particles are detected by MCP (Bilski, 2006). Dosimetry with LiF based phosphors outside the linearity limit (using experimentally determined non-linearity correction functions) was so far possible only up to 1 kGy, which is the level of saturation of the TL main peak for both materials. The high-dose high-temperature TL emission of MCP recently observed at the IFJ enabled dose measurement of up to 1 MGy (Obryk et al., 2011a), which is impossible with MTS. High-dose features of both phosphors have been determined experimentally since the discovery of TL peak 'B' of MCPs and most of them seem to be significantly different.

**2.2. High-dose experiments**

Until now TL emission glow-curves of highly irradiated LiF:Mg,Cu,P and LiF:Mg,Ti detectors have been studied with radiation qualities of a relatively broad LET range: gammas ($^{60}$Co), electrons (6 and 10 MeV), protons (25 MeV and 24 GeV/c), also after exposures to high thermal neutron fluences (up to $3\times10^{15}$ n/cm$^2$), alpha particles (in the range $10^7$–$10^{11}$ particles/cm$^2$), low energy heavy ions ($10^7$–$10^9$ particles/cm$^2$) and in high energy mixed field which consisted of charged hadrons, muons, neutrons as well as photons and electrons with energy spectrum extending from fractions of eV to several hundreds of GeV (up to $10^{15}$ HEH/cm$^2$). TL emission spectra of highly irradiated detectors of both materials, their photoluminescence (PL), optically stimulated luminescence (OSL) and the preliminary characterization of their electron paramagnetic resonance (EPR) signals after high and ultra-high doses have been also studied. In addition thermally- and radiation-induced sensitivity loss and recovery of detectors have been investigated. The data on high-dose experiments with LiF based detectors are summarized in Table 1 together with references to their results, which are described in detail in the next section.

Table 1. Qualities, energies and dose ranges of radiation used for tests of high-dose high-temperature emission of LiF based detectors so far.

| Radiation type | Radiation energy | Dose/Fluence range | Reference |
| --- | --- | --- | --- |
| Gamma | 1.25 MeV | 1 Gy – 1.2 MGy | Bilski et al., 2007b, 2008b; Obryk et al., 2009; Obryk, 2010; Khoury et al., 2011; Gieszczyk et al., 2013b; 2013c |
| Electron | 6 MeV, 10 MeV | 5 kGy - 1 MGy | Bilski et al., 2010; Obryk, 2010; Mrozik et al., 2014 |
| Proton | 25 MeV, 24 GeV/c | 1 Gy - 1 MGy | Obryk et al., 2009, 2010; Obryk, 2010 |
| Neutron | Thermal & epithermal | $3\times10^{11}$ - $3\times10^{15}$ n/cm$^2$ | Obryk, 2010; Obryk et al. 2011b |
| Alpha-particle | 5.5 MeV | $1\times10^7$ - $1\times10^{11}$ α/cm$^2$ | Olko et al., 2011; Gieszczyk et al., 2012 |
| Low energy heavy ion | 5.0 to 9.3 MeV/n | $10^5$-$10^9$ particles/cm$^2$ | Gieszczyk et al., 2013a; Gieszczyk et al., 2014 |
| Mixed field | >20 MeV, HEH | Up to $10^{15}$ HEH/cm$^2$ | Obryk et al., 2008; Obryk, 2010; Mala et al., 2014 |

**3. Characteristics of highly irradiated LiF:Mg,Cu,P and LiF:Mg,Ti phosphors**

The basic difference between the typical TL low-dose glow-curve of LiF:Mg,Cu,P, structure of which doesn't change up to 1 kGy, and high-dose high-temperature glow-curve of this phosphor, with a well separated TL peak 'B', is shown in Fig. 1.

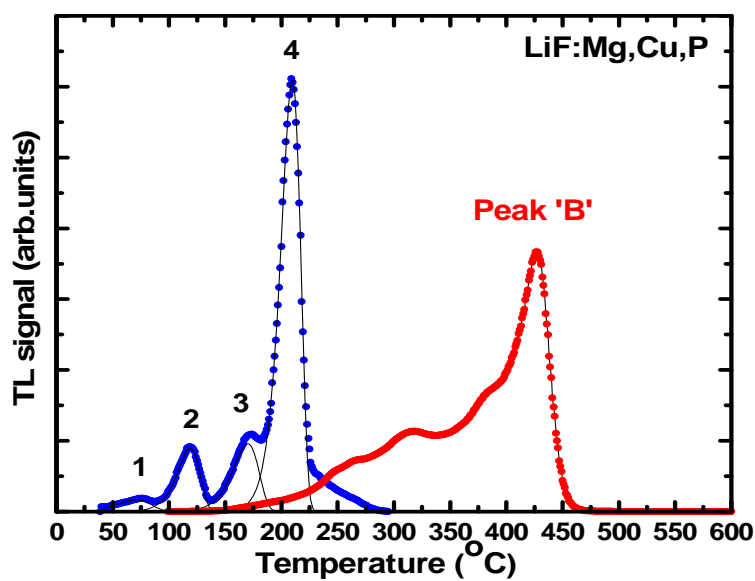

Fig. 1. Comparison of the high-dose high-temperature TL glow-curve of LiF:Mg,Cu,P with a well separated TL peak 'B' (100 kGy – red, with TL intensity damped by a filter) with typical low-dose glow-curve of this phosphor (1.5 mGy - blue).

### 3.1. TL glow-curves for different radiation types

All the TL studies' results showed the presence of the high-dose high-temperature peak 'B' in the glow-curves of LiF:Mg,Cu,P detectors exposed to radiation doses higher than about 30 kGy, following significant changes of their shape for lower doses starting from a few kGy (see Fig. 2 and references quoted in Table 1). Also a distinctive shift of the peak 'B' position towards higher temperatures with increasing dose, in contradiction to standard TL models (Bos, 2006; Chen and McKeever, 1997; McKeever, 1985), was observed for all radiation types (see Fig. 2 for gammas, Bilski et al., 2010 for electrons, Obryk et al., 2010 for protons, Obryk et al., 2011b for neutrons). On the other hand, the situation for LiF:Mg,Ti glow-curves is not consistent across different radiation types. At high doses, up to about 500 kGy, most of the TL signal is emitted below 400°C, but high temperature peaks grow with dose together with the main peak up to about 10 kGy when its saturation starts and high temperature peaks become dominant (see Fig. 3). However, for doses higher than about 100 kGy different glow-curve behaviour was observed for various radiation types studied. At ultra-high doses of gammas (see Khoury et al., 2011) but also after very high thermal neutron fluences (Obryk, 2010) TL emission shifts towards higher temperatures while its amplitude lowers. However, after electron and proton irradiation no significant TL emission is visible above 400°C for doses up to about 1 MGy as presented in Fig. 4 (Bilski et al., 2010; Obryk et al., 2010). For all radiation types studied (including high energy mixed field, see Obryk, 2010), a small residual peak is present in the MTS glow-curves close to 500°C (for gammas and protons see Fig. 3 and Fig. 4, respectively) which behaves differently with the dose than the peak 'B', i.e. starts growing at about 10 kGy, reaching rather flat maximum in the region of doses around 100 kGy, then decreases steeply (Bilski et al., 2010; Obryk, 2010).

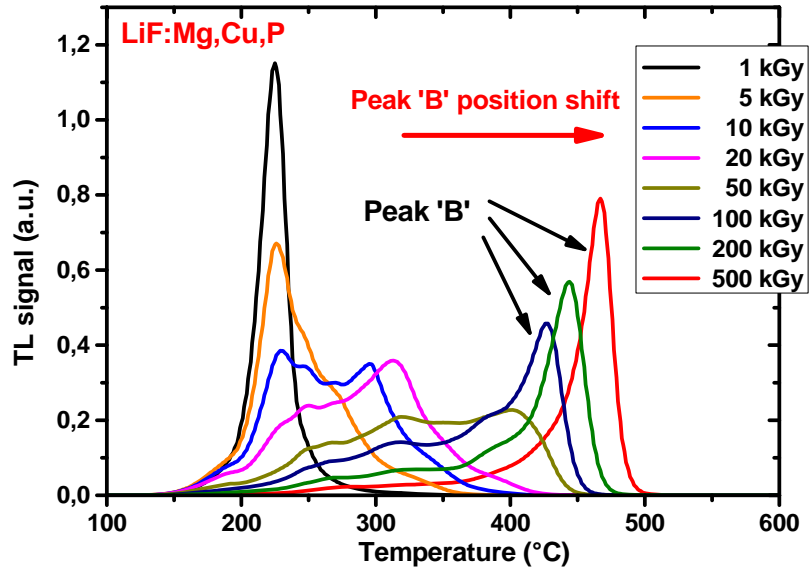

Fig. 2. LiF:Mg,Cu,P glow-curves resulting from gamma irradiation ($^{60}$Co source at KAERI) for the dose range 1 kGy-500 kGy (Obryk et al., 2013).

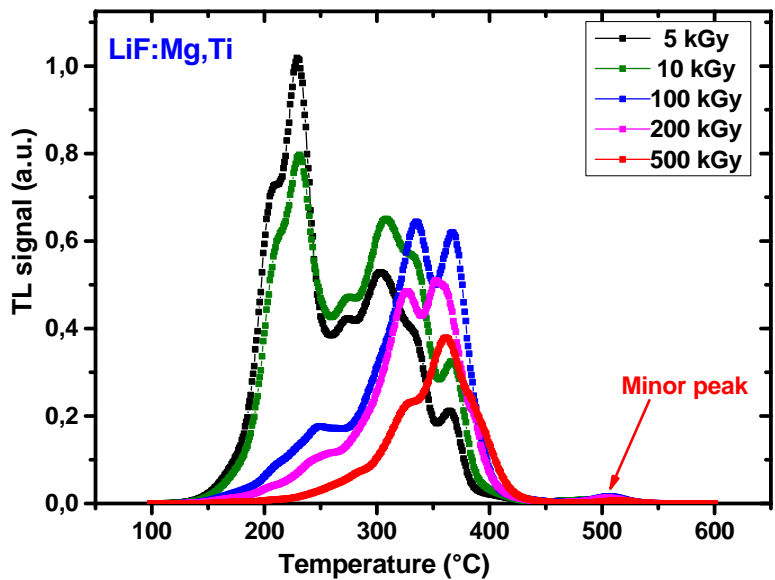

Fig. 3. LiF:Mg,Ti glow-curves resulting from gamma irradiation ($^{60}$Co source at KAERI) for the dose range 1 kGy-500 kGy.

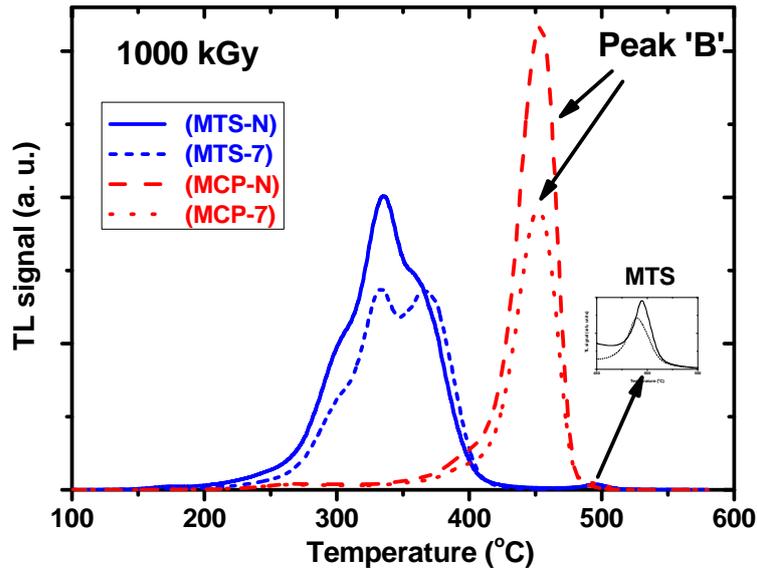

Fig 4. Glow-curves of MTS-N/7 and MCP-N/7 detectors for 1 MGy dose of 24 GeV/c protons (Obryk et al., 2010).

The TL peak 'B' is a distinctive feature of the polycrystalline LiF:Mg,Cu,P samples produced with sintering method and we also observed its presence in the TL emission of highly irradiated LiF:Mg,Cu,P powder samples; however, we found that it did not occur in the TL emission of a single crystal produced from this powder and exposed to 50 kGy gammas.

It has also been observed that two month fading of the TL peak 'B' (Obryk,2010) does not differ significantly from fading of basic peaks of LiF:Mg,Cu,P detectors reported by Ptaszkiewicz (2007). Our recent data, which are still under detailed evaluation, also indicate that the TL peak 'B' intensity was not altered substantially eight years after exposure, on the contrary it even seems as amplification of the TL peak 'B' occurred, and only low temperature peaks were suppressed.

**3.2. TL emission spectra**

It was observed that above 20 kGy TL emission spectrum changes shifting towards longer wavelength with increasing dose (Mandowska et al., 2010). This behaviour was observed for MCP detectors while not for MTS; we have observed that for MTS detectors after 80 kGy gamma dose, spectra is only just a bit more stretched towards longer wavelength. However, all these spectral analysis have been obtained with detectors heated only up to 350°C due to technical limitations. Hence the temperature of the TL peak 'B' emission has not been reached within these studies and this peak is not present in these results.

Results obtained from the experiment performed at the Delft University of Technology, Delft, The Netherlands, with heating of the MCP detectors up to 550°C have shown spectral characteristic of the TL emission at temperatures where the TL peak 'B' is present (Gieszczyk et al., 2013b; 2013c). It was observed that emission spectrum of the TL peak 'B' is at 300-400 nm, so at the same band as the TL emission of the main dosimetric peak (peak 4) of this material. This suggests that no new recombination centres are created at high doses, only the population of the known centres depends on the dose.

Due to significant presence of the long wavelength band in the TL emission of MCPs for doses in several kGy region, one has to be careful with spectral characteristics of the TL equipment used when experimenting on detectors exposed to high doses. Due to the difference in short and long wavelength bands' contribution to the TL spectrum of highly irradiated MCPs in the range from 1 kGy to 1 MGy, one has to be aware of PMT spectral characteristic and filters' transmittance, in addition to spectral transmittance of high dose filters used in order to damp intensive TL signal before it reaches PMT. Not paying attention to this makes it possible to create artificial differences in the TL glow-curves obtained, especially in several kGy dose region (Bilski et al., 2014a).

**3.3. Colouring of detectors**

The effect of colouring the TL detectors based on LiF after high-dose exposure has been observed. At doses of a few kGy gradual yellowing of the detectors starts, as the dose increases the colour darkens, reaching dark brown for doses of about 1 MGy. This effect was observed in all high-dose experiments carried out. However, colouring of detectors depends on the quality of the radiation used and irradiated material type (see Fig. 5).

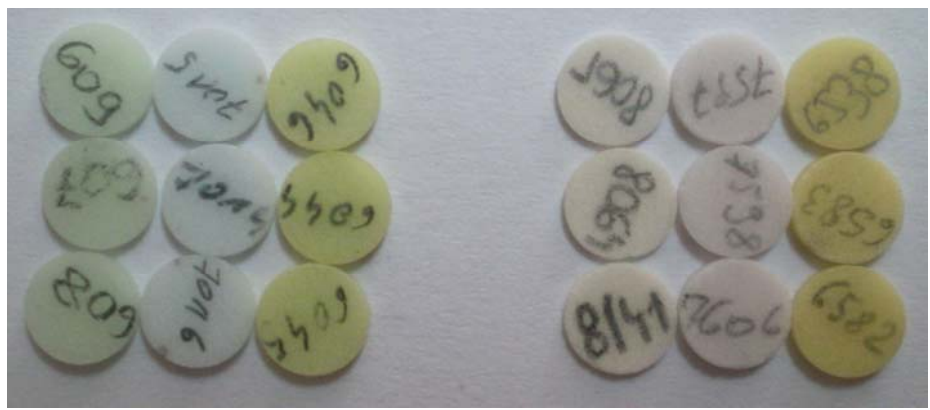

Fig. 5. Photograph illustrating the different colours of different LiF based detector types after high dose (about 4 kGy) exposure to mixed field with thermal neutron presence (at IRRAD6 PS at CERN): LiF:Mg,Cu,P detectors on the left, LiF:Mg,Ti on the right, $^{nat}$Li pellets in the first column, $^7$Li enriched in the middle and $^6$Li enriched in the third column of each of two presented samples.

This phenomenon relies on the colour centres formation under the influence of ionizing radiation. These centres alone or in agglomerations with other defects constitute centres of luminescence in many TL materials (McKeever, 1985; Townsend and Kelly, 1973). However, after being heated up to about 600°C during readout all the detectors regained their original colour (see Fig. 6). This seems to be caused by the annihilation of the F centres with the H centres, which removes lattice defects (McKeever, 1985).

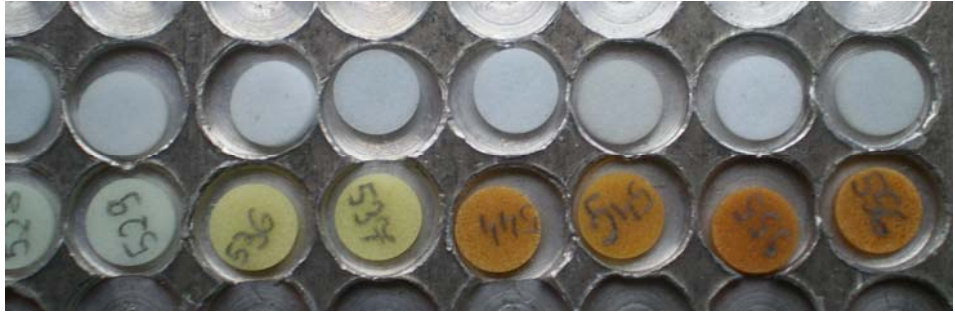

Fig. 6. Photograph illustrating the recovery of the original colour by the LiF:Mg,Cu,P detectors irradiated with high-doses of 24 GeV/c protons (Obryk, 2010) in the process of reading up to 600°C (dose is growing from left to right, in the upper row are detectors after readout process).

Dose-dependent colour of LiF detectors carries great potential for high-level dosimetry with proper instrumentation used for analysis of this effect. Studies on behaviour of colour centres in highly irradiated LiF based detectors by photoluminescence emission and EPR spectroscopy could also add some information to the understanding of the TL peak 'B' formation mechanism.

### 3.4. Photoluminescence (PL)

A difference between the PL signals of highly irradiated LiF:Mg,Cu,P and LiF:Mg,Ti detectors was observed. The PL spectra obtained after 405 nm wavelength excitation show two wide emission bands for both highly irradiated phosphors, more intensive one at about 525 nm and minor one at about 640 nm. However, while for both phosphors PL is at the similar intensity level for doses up to tens of kGy, intensity of PL for both bands is much higher for LiF:Mg,Cu,P than for LiF:Mg,Ti at higher doses. It is worth to mention that the time gap between exposure and measurement was more than two month and these PL spectral studies are under verification and details will be published separately.

Mrozik et al. (2014) reported linearity of the PL dose response for LiF:Mg,Cu,P up to about 50 kGy while for LiF:Mg,Ti linearity ranges only up to a few kGy and then it becomes sublinear for both phosphors. Saturation of the PL signal was found for both phosphors for doses higher than about 100 kGy. The PL signal gradually decreases to background level after heating to temperatures in the range of 150 C-270 C; simultaneously detectors lose colouring with increasing temperature, reaching the original colour at about 270 C, at which the TL peak 'B' emission did not start yet. Hence the PL signal rather than TL seems to be connected with the colour centres, and it suggests that two different mechanisms are responsible for both phenomena.

### 3.5. Optically stimulated luminescence (OSL)

Bilski et al. (2014b) observed that both LiF:Mg,Cu,P and LiF:Mg,Ti exhibit OSL properties in the low dose range (below 1 Gy). They identified quite high LiF:Mg,Cu,P OSL signal as connected directly to the TL peak 2, while for LiF:Mg,Ti situation is more complex according to them and suggests a complex structure of the relevant trapping sites. Pre-heat above 120°C significantly reduces OSL intensity, but for LiF:Mg,Cu,P even after pre-heat at 240°C a significant OSL signal was still observed.

Comprehensive measurements of OSL signal of highly irradiated LiF based phosphors have been also performed very recently and will be published separately. However, from their first evaluation it seems that LiF:Mg,Ti OSL in the range of kGy is showing no significant differences, while for LiF:Mg,Cu,P the highest OSL signal have been observed for the range of tens of kGy, then it decreases for 100 kGy reaching back the level of the OSL signal observed after 1 kGy exposure of this material.

### 3.6. Electron Paramagnetic Resonance (EPR) spectra

Variations of the intensity of the EPR signal with increasing dose were analysed for both phosphors. For highly irradiated MTS detectors two EPR signals were identified at room temperature (Khoury et al., 2011). The first one (at g=2.0089) increases with the increasing dose until 20 kGy and then decreases, saturating at 300 kGy. The second EPR peak (g=1.9863) appears at 100 kGy, grows up to 700 kGy and then remains constant up to 1200 kGy, and this appears to be associated with high-temperature TL peaks observed for this material at this range of doses. The first EPR signal seems to correspond to a paramagnetic centre with electron deficiency while the second one to a centre with electron excess.

However, we observed that two EPR signals identified for highly irradiated MCP detectors, differently than for MTS, do not present a significant variation with the dose. We suspected that the defect centre responsible for the TL peak 'B' emission has no paramagnetic proprieties or cannot be detected at room temperature. In order to obtain more accurate results for LiF:Mg,Cu,P we have performed EPR studies of the powder samples of this phosphor. We have identified three EPR signals (at g=2.1940; 2.0040; 1.9850) with different intensities which behave differently with dose but detailed analysis of these results will be published separately.

### 3.7. Sensitivity damage and recovery

So far all high-dose experiments and measurements have been performed on fresh samples and each detector was used only once, because LiF:Mg,Cu,P detectors lose their sensitivity to a large extent as a result of such exposures (Cai et al., 1994; Meijvogel and Bos, 1995). However, Bilski et al. (2008a) reported that LiF:Mg,Cu,P is more resistant to radiation-induced sensitivity damage than LiF:Mg,Ti. They observed no decrease of MCP sensitivity after pre-doses of up to 100 Gy, while for MTS sensitivity loss after doses higher than 100 Gy appears to be irreversible since even the longest applied annealing (21 hours) at 400°C did not recover their original sensitivity.

Nevertheless, well known MCPs' feature is their thermally induced sensitivity loss when heated beyond about 270°C (e.g. Oster et al., 1993; Tang et al., 2000), hence following high-dose high-temperature measurements this is entangled with radiation-induced sensitivity loss. These effects were thoroughly studied very recently and it had been proven that thermally and radiation-induced sensitivity loss of LiF:Mg,Cu,P is fully reversible up to pre-dose of 100 kGy, using high-temperature annealing procedure developed by Obryk et al. (2013), which takes less than an hour. This indicates that high temperature and high doses of radiation do not cause irreversible changes in the MCP material.

## 4. Summary and Discussion

We have shown that very many experimental data have been obtained by a few research groups, we believe that we referred to most of them. However, despite all these experimental efforts, understanding of the mechanism of high-dose high-temperature TL peak 'B' emission is still very limited. Some hypotheses have been made; however, they were not verified in detail (e.g. Gieszczyk et al., 2013b). The results of X-ray diffraction have found similar structural parameters of not irradiated LiF:Mg,Cu,P detector, irradiated with 1 MGy, and readout detector (Gieszczyk, 2013). It allows to conclude that no significant changes in the structure of this material (in particular no phase transition) are connected with high-dose high-temperature TL emission.

Recently Chen and Pagonis (2014) achieved interesting results in their comprehensive simulations consisting of the numerical solution of sets of coupled nonlinear equations of the TL model. Of special interest for high-dose dosimetry among their results are a shift of the position of the glow-peak to higher temperature and unusually high values of activation energy and frequency factor, which are in agreement with unpublished results of glow-curve deconvolution of MCPs' TL emission after 24 GeV/c proton exposure in the dose range 10 kGy – 1 MGy, obtained by Kitis and Obryk. The analytical solutions of these sets of equations based on simplified assumptions were unable to provide such results.

It was not only proven that LiF:Mg,Cu,P phosphor is a very promising high-level dosimetry material but also an ultra-high dose range dosimetry method has been proposed and is applied in practice for a few years (e.g. at CERN, see Mala et al., 2014, and at JET, see Obryk et al., 2014) based on its unprecedented luminescent features. Following this research high dose features of some other luminescent materials have been observed (e.g. Kortov and Ustyantsev, 2013). This all allows us to conclude that the luminescent methods have a great potential for serving high-level dosimetry (Olko, 2010).

In connection with the observed PL emission of LiF:Mg,Cu,P, whose intensity also grows with dose as TL peak 'B' intensity, it is possible to propose a combined PL/TL high dose dosimetry method, where at first PL signal would be measured for MCP detector, which does not affect its TL emission, and then TL readout of the same detector would be performed. This would allow a significant improvement of the accuracy of high-level TL dosimetry with LiF:Mg,Cu,P.

Finally, we have concluded that dopants are playing a crucial role in high-dose features of lithium fluoride based phosphors. Our next approach is to study the response of these phosphors with modified dopants' concentration to high and ultra-high radiation doses with the aim of gaining more knowledge about the nature of the trapping and recombination centres responsible for the peak 'B' emission, verifying in particular a hypothesis that some clustering processes are responsible for this and other features of highly irradiated lithium fluoride phosphors.

**Acknowledgements**

Barbara Obryk sincerely thanks the Scientific Advisory Committee of the 17th Solid State Dosimetry Conference for the invitation to present a talk and financial support of participation. This work was partly supported by a research project from the National Science Centre, Poland (Contract No. 2013/09/D/ST2/03718).